# X-ray Linear Dichroic Tomography of Crystallographic and Topological Defects

**Author List:**

Andreas Apseros*,[1,2] Valerio Scagnoli*,[1,2] Mirko Holler,[2] Manuel Guizar-Sicairos,[2,3] Zirui Gao,[1,2,4] Christian Appel,[2] Laura J. Heyderman,[1,2] Claire Donnelly*,[5,6] and Johannes Ihli[2,7]

**Affiliation:**

[1] ETH Zürich, 8093 Zürich, Switzerland

[2] Paul Scherrer Institute, 5232 Villigen PSI, Switzerland

[3] EPF Lausanne, 1015-Lausanne, Switzerland

[4] Brookhaven National Laboratory, Upton NY 11973-5000, United States of America

[5] Max Planck Institute for Chemical Physics of Solids, 01187 Dresden, Germany

[6] International Institute for Sustainability with Knotted Chiral Meta Matter (WPI-SKCM2), Hiroshima University, Hiroshima 739-8526, Japan

[7] University of Oxford, OX2 6NN Oxford, United Kingdom

*Correspondence and requests for materials should be addressed to andreas.apseros@psi.ch, valerio.scagnoli@psi.ch , claire.donnelly@cpfs.mpg.de

**Keywords:** • Ptychography • Tomography • Linear Dichroism • X-ray Spectroscopy • Microstructure • Crystalline Materials

**"One-Sentence" Summary:**

The properties of functional materials are determined by their microstructure – i.e., the distribution and orientation of crystalline grains and the defects within them on the micro- and nanoscale. Here, we introduce ptychographic X-ray linear dichroic orientation tomography, a non-destructive technique that provides intra- and inter-granular characterisation of materials with nanoscale spatial resolution within extended crystalline and non-crystalline samples.

**Abstract**

The functionality of materials is determined by their composition[1–4] and microstructure, that is, the distribution and orientation of crystalline grains, grain boundaries and the defects within them[5,6]. The characterisation of the material's microstructure is therefore critical for materials applications such as catalysis, energy storage and buildings. Until now, characterization techniques that map the distribution of grains, their orientation, and the presence of defects have either been limited to surface investigations, to spatial resolutions of a few hundred nanometres, or to systems of thickness around one hundred nanometres, thus requiring destructive sample preparation for measurements and preventing the study of system-representative volumes or the investigation of materials under operational conditions[7–15]. Here, we present X-ray linear dichroic orientation tomography, a quantitative, non-invasive technique that allows for an intra- and inter-granular characterisation of extended polycrystalline and amorphous materials in three dimensions (3D). We present the detailed characterisation of a polycrystalline sample of vanadium pentoxide ($V_2O_5$), a key catalyst in the production of sulfuric acid[16,17]. In addition to determining the nanoscale composition, we map the crystal orientation throughout the polycrystalline sample with 73 nm spatial resolution. We identify grains, as well as twist, tilt, and twin grain boundaries. We further observe the creation and annihilation of topological defects promoted by the presence of volume crystallographic defects in 3D. Our method's non-destructive and spectroscopic nature opens the door to in-operando combined chemical and microstructural investigations[11,18] of functional materials, including energy and mechanical materials in existing industries, as well as quantum materials for future technologies[19].

**Main:**

Materials properties and functionality depend on the material composition and microstructure. The application-driven manufacturing of materials therefore requires an understanding of the underlying structure and composition, often at the nanoscale, as well as their link to the functionality. This is of paramount importance across many fields including catalysis[3], energy storage and conversion[20,21], and permanent magnets[22]. While compositional mapping down to the nanoscale, even across extended volumes, can be realised for example, with high spatial resolution mapping of the electron density[11], characterisation of the distribution, type, and topology of crystallographic defects, and their dynamic behaviour under external stimuli, such as temperature, pressure, or electromagnetic fields, remains a major challenge. To map the microstructure, including crystal grains and associated defects, and determine their structure-functionality relationship, 3D nanoscale mapping of the orientation within extended systems is key.

So far, mapping of the local crystal orientation has been possible with electron-based techniques such as Transmission Electron Microscopy[7,8] (TEM) and Electron Back-Scatter Diffraction[13] (EBSD), which can achieve sub-10 nm spatial resolution with planar measurements. However, as these measurements are limited to materials of thickness on the order of 100 nm, destructive serial sectioning methods are required to acquire full 3D orientation maps. The non-destructive imaging of micrometre-thick materials has been addressed with Diffraction Contrast Tomography[15] (DCT) but this technique is limited to the size of the X-ray probe, with a spatial resolution of hundreds of nanometres, and can only be used to characterise crystalline samples.

Here, we present the 3D mapping of nanoscale grains, grain boundaries, and topological defects in polycrystalline $V_2O_5$ enabled by X-ray linear dichroic orientation tomography (XL-DOT). By acquiring high spatial resolution synchrotron X-ray ptychographic projections of the linear dichroism, which is sensitive to the local crystal orientation, at different sample orientations and implementing a tailored reconstruction algorithm, we recover the 3D map of the local crystalline structure with nanoscale spatial resolution within a 6 µm size sample. The resulting tomogram provides the local orientation of the $c$-axis of $V_2O_5$, allowing for the nanoscale segmentation of grains within the sample that are otherwise indistinguishable in the electron density contrast tomogram. This quantitative, non-invasive, and simultaneous intra- and inter-granular characterisation of extended polycrystalline and amorphous materials in 3D on the nanoscale is a key step towards the non-destructive operando imaging of material microstructure.

We demonstrate the capabilities of XL-DOT by mapping the underlying microstructure of a nanoporous $V_2O_5$, pillar prepared with high-temperature sintering, which is 6 µm in diameter, and consists of randomly orientated, oxygen vacancy defect-rich, orthorhombic α-$V_2O_5$ crystals with an average grain size greater than 300 nm (Figure S1-3). The effectiveness of $V_2O_5$ as a heterogeneous catalyst in the chemical industry is closely related to its microstructure[16,17]. Therefore, a detailed mapping the grain structure on the nanoscale will provide a route to understanding and optimising its properties.

To obtain a map of the grain orientation and defects in the polycrystalline sample, we determine the local orientation of the $V_2O_5$ crystallographic $c$-axis in 3D by performing XL-DOT. The experimental setup is shown schematically in Figure 1a, along with an illustration explaining linear dichroic contrast in Figure 1b. As opposed to conventional tomography, where the acquisition of projections around a single tomographic rotation axis is sufficient to describe a scalar component, orientation tomography – similar to vector tomography[23] – requires the acquisition of projections around multiple tomographic rotation axes using a component-sensitive probe to access and characterise each of the three components of the orientation[23,24]. Here, to gain sensitivity to the local $c$-axis orientation of the unit cell, we harness X-ray linear dichroism, where the transmission function of the sample depends on the relative angle between the incident X-ray beam polarization and the crystallographic $c$-axis. High spatial resolution projections with strong contrast were acquired by probing the linear dichroic component, and thus the local microstructure of the $V_2O_5$, with resonant dichroic X-ray ptychography[12,25,26]. For this, the energy of the X-rays was tuned to the pre-edge of the V K edge (5.468 keV). Using ptychography, phase projections were obtained, enabling the detection of the relatively weak linear dichroism signal[27,28] (Figure S4-6) as shown in Figures 1c, d. Tomograms, consisting of 280 phase projections measured at regular angular intervals over 180º, were obtained for four sample tilts (see four panels to the right of Figure 1a), with both linear horizontal (LH) and linear vertical (LV) polarizations, resulting in a total of 2,240 phase projections. The tomographic projections were then aligned with high precision[29] and the 3D $c$-axis orientation tomogram – or map – was reconstructed using a gradient-based iterative reconstruction algorithm as described in the Methods and shown schematically in Figure S7. Within the resulting $c$-axis orientation tomogram, we can resolve features of 40 nm in width, with a calculated average spatial resolution of 73 nm throughout the sample (see Methods, Figure S8).

As well as the $c$-axis orientation tomogram acquired on resonance, a high spatial resolution tomogram of the electron density was obtained with the X-rays tuned to an energy far from the absorption edge of 5.4 keV. There are no dichroic contributions to this non-resonant measurement, which provides a quantitative electron density tomogram with a spatial resolution of 44 nm, allowing for local compositional analysis[30] (Figure 2a-c & S9-10).

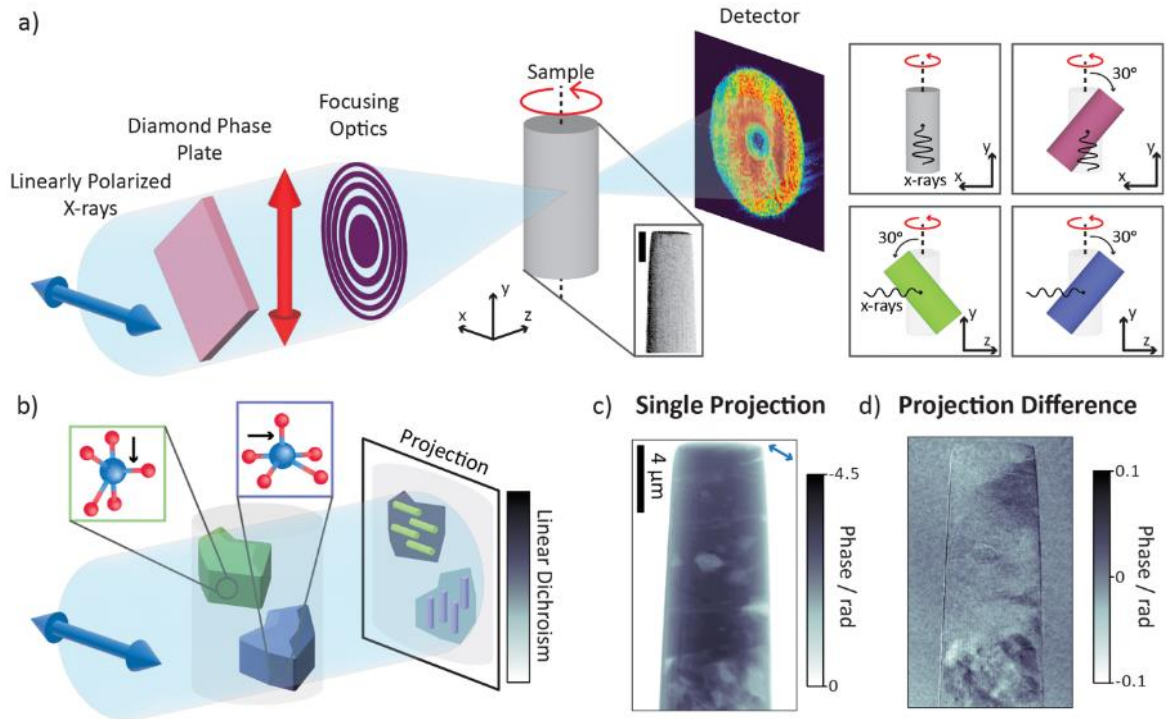

**Figure 1: X-ray Linear Dichroic Orientation Tomography (XL-DOT) Experimental Setup**. a) Illustration of the XL-DOT setup. Tomographic projections using linear horizontal (LH ↔) and linear vertical (LV ↕) polarized X-rays were acquired through synchrotron X-ray ptychography. The sample was illuminated with X-rays using a Fresnel zone plate with engineered aberrations.[31]. The sample was scanned in the x-y plane, with mutually overlapping illuminations. At each scanning point, a far-field diffraction pattern was acquired. A ptychographic reconstruction of these patterns returns phase and amplitude projections. Scanning electron micrograph (SEM) of the studied cylindrical polycrystalline $V_2O_5$ sample is shown in the inset. A diamond phase plate was used to change the sample incident X-ray polarisation. To probe and later reconstruct all three components of the crystallographic *c*-axis orientation, tomographic data was collected at four sample tilts, represented by the coloured cylinders in the four panels to the right, each with LH and LV polarized illumination. The reference orientation is shown as a half-shaded grey pillar in all panels. b) Illustration of X-ray linear dichroism, showcased with the imaging of two grains of $V_2O_5$ (in green and blue). The transmission function of $V_2O_5$ under resonance conditions depends on the relative orientation of the Vanadyl bond (indicated with an arrow), which is parallel to the crystallographic *c*-axis, and the linear polarisation state of the illumination. Overlaid on the example projection are coloured rods that represent the orientation of the crystallographic *c*-axis in the grains of the same colour. c, d) Phase contrast projection of the sample acquired with LH polarisation, whose relative orientation is shown by the double headed arrow and e) the X-ray Linear Dichroic (XLD) difference between projections acquired with LH and LV, illustrating electronic and linear dichroic phase contrast contributions, respectively. Scale bars are 4 μm.

We first examine the off-resonance tomogram that provides a quantitative, high spatial resolution mapping of the sample's electron density, allowing us to identify the presence of different materials within the sample. The electron density tomogram and a virtual cut are plotted in Figure 2a and 2b, respectively. We find that, alongside solid regions corresponding to the electron density of $V_2O_5$, both nanoporous regions containing air and polystyrene, a residue of the synthesis process, are present, as identified quantitatively in the histogram in Figure 2c. In the $V_2O_5$ regions, we observe two main forms. In the upper half of the sample (Figure 2a), a rather large, continuous volume of $V_2O_5$ is observed. In the lower half of the sample, one can identify nanosized features whose orthorhombic shape reflects the favoured crystal phase of $V_2O_5$. Although the morphology seen in the electron density tomogram indicates the presence of grain-like features such as the elongated grain highlighted in Figure 2b, without knowledge of the crystallographic orientation, it is not possible to determine whether these are indeed single crystal grains or volumes of polycrystalline material.

The single crystal and polycrystalline volumes can be distinguished by considering the 3D $c$-axis orientation map obtained with XL-DOT. Plotting the $c$-axis orientation map, shown in Figure 2d, 2e, we immediately see a rich, complex structure with a large variety of local crystal orientations. We first can confirm that the elongated grain identified in the electron density map in Figure 2b is indeed a single crystallite within our resolution (dashed black box in Figure 2e, shown in full in Figure 2j). Another example of a grain is shown in Figure 2g, and its location is marked with a dashed red box in Figure 2e. However, we also observe regions that turn out to be polycrystalline, exhibiting abrupt changes in the crystal orientation that could not be easily distinguished in the electron density tomogram. One example is marked in the full red box in Figure 2e and shown in full in Figure 2f. Animated views of the morphology, as well as slices comparing the electron density and orientation tomogram are presented in Movies S1&S2, respectively.

The high spatial resolution map of the $c$-axis orientation allows us to detect grain boundaries associated with an angular mismatch down to 10˚ (Figure S11-12), a precision that, in principle, allows for the detection of more than 95% of grain boundaries in randomly oriented samples[32]. We use the orientation tomogram to segment grains of distinct crystallographic orientations. Each separated volume represents a different $c$-axis orientation, as shown in Figure 2h.

The successful segmentation of the crystallites, shown in Figure 2h, also allows us to extract quantitative information about the morphology of the grains throughout the sample. The correlation between the size and shape of the grains is shown in the histogram in Figure 2i. A strong negative correlation is observed between the size and shape of the particle, with larger grains exhibiting more elongated geometries. This is consistent with the orthorhombic morphology that is favoured by the growing crystal. With this segmentation, it is possible to observe that larger, more elongated grains (average sphericity = 0.26, example shown in Figure 2j) up to 1.5 µm in length occur as free-standing objects in the more porous regions, while the smaller, more spherical grains (average sphericity = 0.3, example shown in Figure 2k) of average length 1.3 µm are clustered in close-packed regions of the sample.

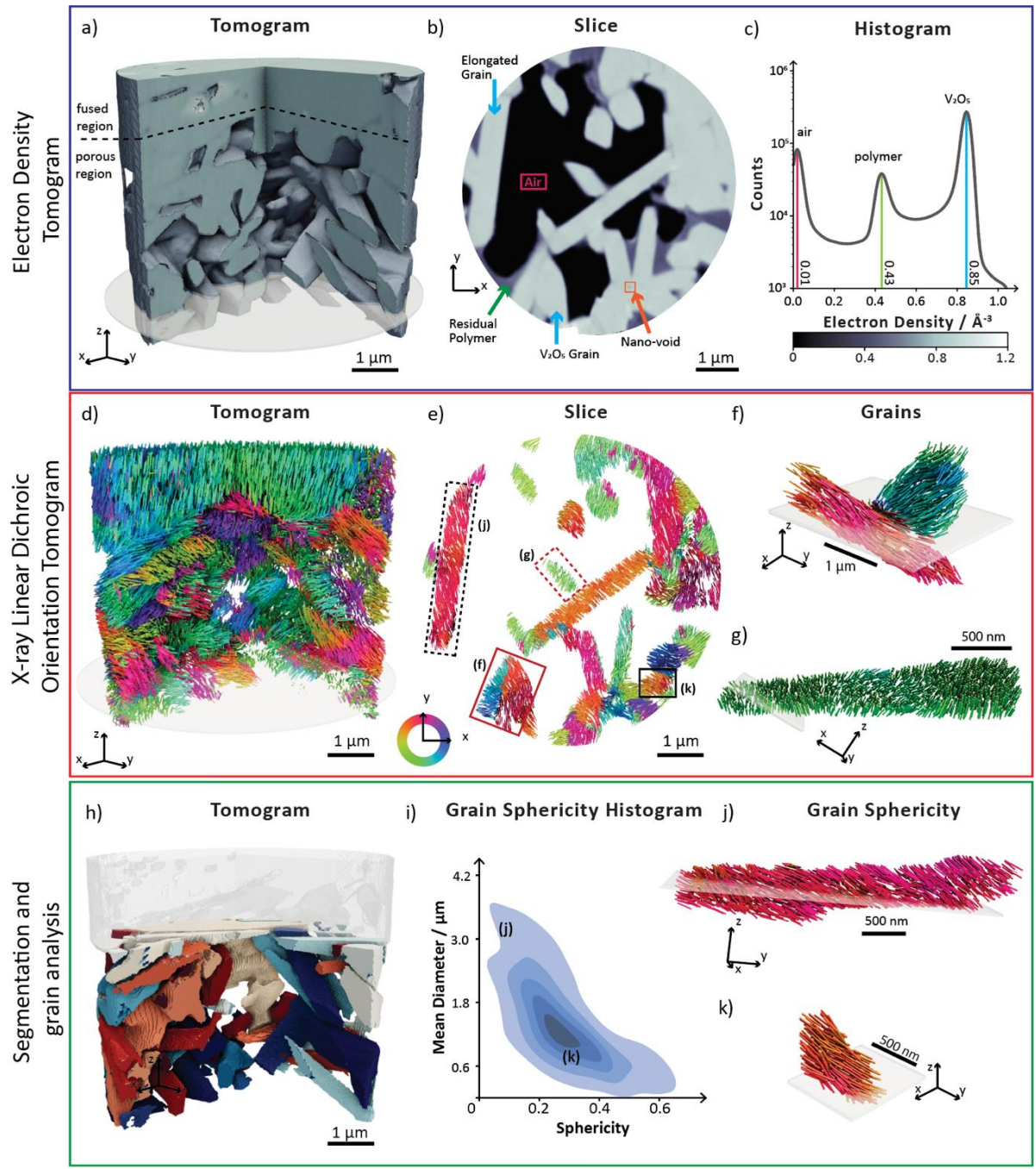

**Figure 2: 3D Mapping the Composition and Microstructure of Materials on the Nanoscale with Linear Dichroic Orientation Tomography (XL-DOT).** (a-c) Electron density tomography data. a, b) Volume rendering and horizontal slice through the tomogram reveal the sample's constituent materials and morphology. c) Histogram of the electron density determined from the tomogram shown in a). The sample components: $V_2O_5$, polystyrene polymer and air, are indicated. (d-k) XL-DOT reconstruction. d, e) Volume rendering and horizontal slice through the *c*-axis orientation tomogram, retrieved through XL-DOT, revealing the sample's microstructure. Coloured rods represent the alignment of the crystallographic c-axis in each voxel and are coloured according to their xy-plane *c*-axis orientation. f, g) Volume rendering of selected regions of uniform electron density, highlighting ability to distinguish polycrystalline regions and single crystals with XL-DOT. h) Volume rendering of the *c*-axis orientation-segmented tomogram, allowing for the identification of the sample's individual $V_2O_5$ grains. i) Quantitative analysis of the grain morphology reveals a strong correlation between the size and sphericity of the grains. j, k) Example of an elongated grain and a smaller grain with

sphericity and mean diameter indicated in i). Further inter-granular material characteristics, including electron density correlations, are given in Figure S13. The locations of the highlighted grains (f, g, j, k) are indicated in (e) within dashed boxes. Transparent planes in (f, g, j, k) show the location where the volumes intersect with the slice in (e).

In addition to the isolation of individual grains, the 3D *c*-axis orientation map allows for the identification of crystallographic defects, including grain boundaries and textures within regions that appear continuous in the electron density. For example, we can detect and differentiate between planar inter- and intra-granular crystallographic defects. In particular, it is possible to identify and map the local configuration of tilt, twist and twin grain boundaries within grains. We first consider a region where the morphology seen in electron density tomogram (see Figure 3a) indicates the meeting of two elongated grains. Within the limit of our spatial resolution, the electron density in the structure is homogeneous, and no boundary can be observed. However, the *c*-axis orientation plot (Figure 3b) reveals a distinct, abrupt grain boundary. The nature of this grain boundary becomes clear from the two grain orientations. In both grains the *c*-axis is oriented at an angle of $30°\pm6°$ to the grain boundary, exhibiting mirror symmetry across the boundary, and thus indicating the presence of a twin defect schematically shown in Figure 3c.

As well as investigating boundaries within free-standing grains, we can also take a closer look at the features that occur within the close-packed region where the electron density appears homogenous within the limit of our spatial resolution. Such a region is shown in Figure 3d where, in the XL-DOT reconstruction image (Figure 3e), also shown schematically in Figure 3f, we can identify four regions with different *c*-axis orientations. In this polycrystalline region, we can identify two types of grain boundaries. In the upper part, going from left to right across the boundary shown in Figure 3g, the crystal orientation rotates from an in-plane (xy-plane) to an out-of-plane orientation, about a rotation axis parallel to the grain boundary normal: a prototypical *twist* grain boundary. In contrast, in the lower part, going from left to right across the boundary shown in Figure 3h, the crystal orientation rotates abruptly from an in-plane (xy-plane) to an out-of-plane orientation about a rotation axis perpendicular to the grain boundary normal: a *tilt* grain boundary. We note that these two examples of tilt and twist boundaries are associated with a high rotation angle of approximately $86°\pm6°$ and $65°\pm6°$, respectively. A low-angle tilt grain boundary is also observed within the left-hand region (circled in Figure 3e) facilitating the transformation of the grain boundary from twist to tilt.

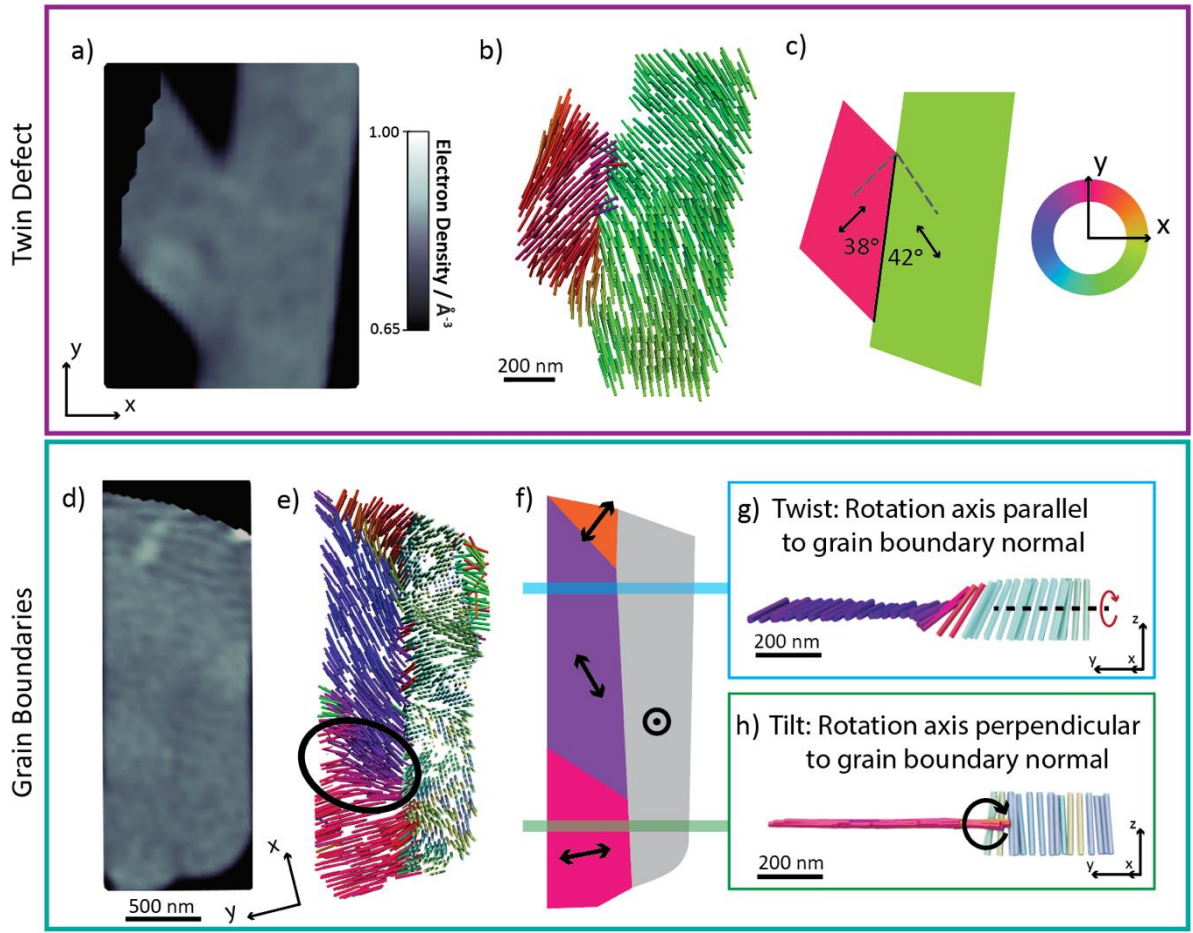

**Figure 3: Crystallographic Defects Detected using Ptychographic X-ray Linear Dichroic Orientation Tomography (XL-DOT).** (a, d) electron density renderings of selected regions displaying a uniform electron density in the $V_2O_5$ sample. (b, e) XL-DOT extracted *c*-axis orientation of the same regions revealing their complex internal microstructure, and (c, f) schematic illustrations of the same region indicating XL-DOT detectable crystallographic defects in these areas. (a-c) Twin grain boundary between two grains, with the *c*-axis orientation displaying mirror symmetry about the abrupt grain boundary. (d-h) Twist and tilt grain boundaries within a region of homogeneous electron density. XL-DOT reveals the presence of four grains with distinct *c*-axis orientations. Between the grains, we observe high angle twist and tilt boundaries, also shown schematically in g, h), as well as a low angle tilt boundary (circled in e). The orientation of the crystallographic *c*-axis is indicated by the colour scheme, with the colour representing the xy-plane orientation. The colours become lighter as the out-of-plane tilt increases. Electron density renderings (a, d) share the same colour scale.

As well as grain orientation, XL-DOT reveals the presence of topological defects, which can subsequently be characterised. When considering the x-y components of the orientation, we observe comet and trefoil defects[33] that also exist in nematic and biological systems[34]. These topological defects are associated with a local topological charge of +1/2 and -1/2 for the comet defect and trefoil defect, respectively. In an infinite system, the overall topological charge must be conserved and thus such defects can only be created or destroyed in pairs, or at surface boundaries.

In Figure 4, we map the spatial evolution of crystalline topological defects in the vicinity of crystallographic volume defects. The layers, shown in Figure 4a, are numbered (i-v) and colour-coded to match the corresponding boxes. In the lowest slice (i, red) a single trefoil defect (T1 in orange) exists in the vicinity of the large void. As we move upwards though the crystal (ii, orange), we observe the trefoil defect T1 shift laterally such that its core is centred on a neighbouring nano-void with the surrounding topological structure maintained. Moving further upwards (iii, green) to a layer above the nano-void, the trefoil defect T1 is again laterally displaced into the crystal. In the next layer (iv, blue), we observe the creation of a new trefoil defect (T2 in green) with a comet defect (C in blue), emerging from the surface of the large void. Since they are of opposite topological charge, the overall topology of the system is conserved. Remarkably, in the next slice (v, purple), the comet defect (in blue) and the original trefoil defect T1, annihilate leaving behind a single trefoil defect T2.

The spatial creation and annihilation of topological defects in the $V_2O_5$ crystal is reminiscent of similar behaviour observed in the dynamics of topological defects in liquid crystals, biological systems and magnets. In particular, the recombination of the comet and trefoil defects is smooth and provides insight into the temporal evolution of the 3D crystal orientation during the high temperature sintering process used to produce the sample.

The structural volume defects play an important role in the evolution of the topological defects. Since these are confined volume defects, they do not provide an infinite boundary, so that the overall topology of the system is conserved. Nevertheless, the interruption of the continuous crystal leads to a strong interaction between nano-voids and the trefoil defects. This can be seen with the pinning of the original trefoil (T1 in orange) at the smaller nano-void (Figure 4ii). The core of the trefoil defect is associated with strong lattice bending or deformation, so its removal through the presence of the nanovoid minimizes the lattice energy. Moreover, the injection of the defect pair from the surface indicates that, although the overall topology must be conserved, the surface promotes the creation of defects in the system.

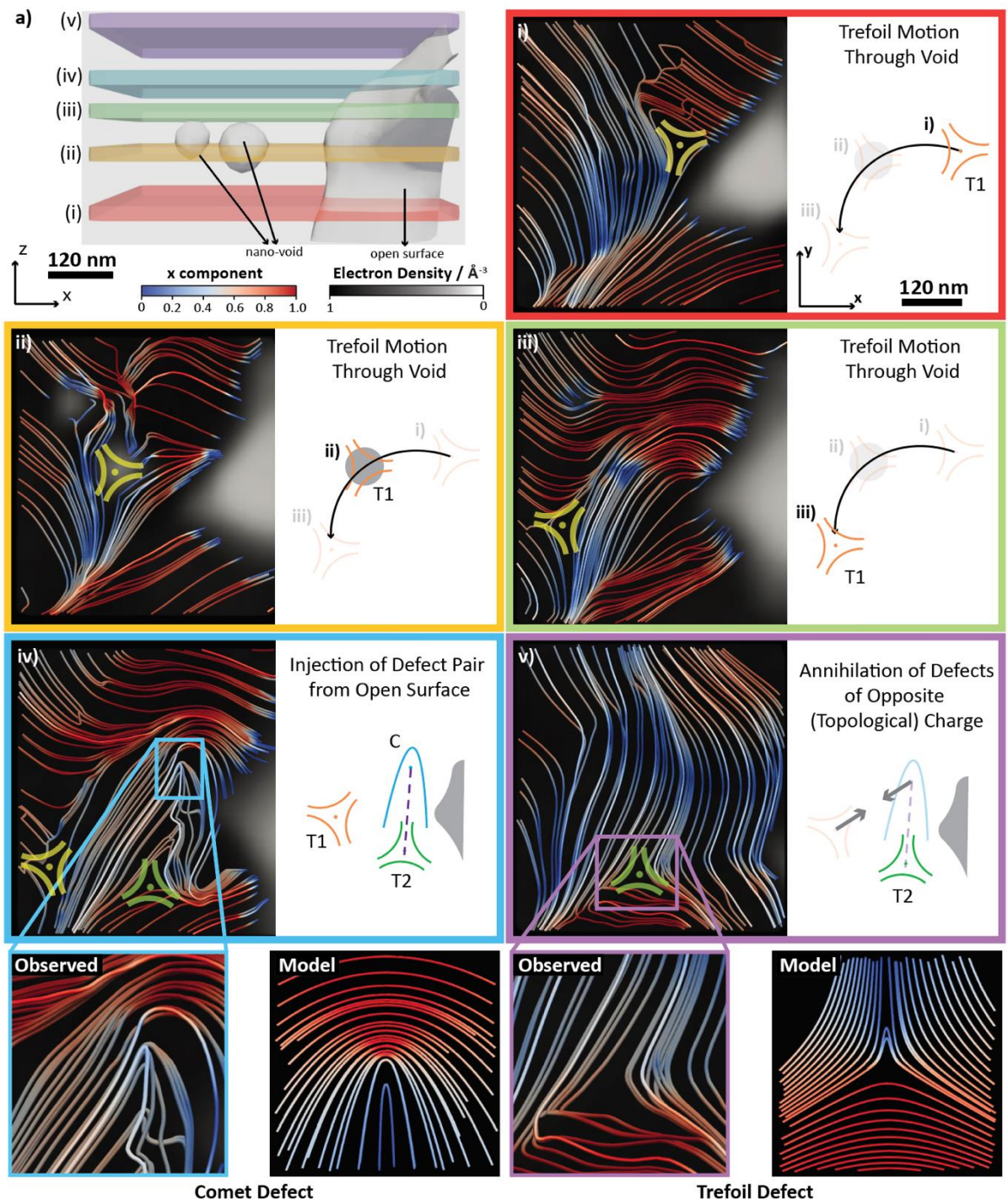

**Figure 4: Crystallographic Volume and Topological Defects.** a) Region of sample in which crystallographic volume defects and topological defects were observed, consisting of an open surface and two closed surface nano-voids, indicated with white isosurfaces. Coloured slabs indicate the height of the layers shown in i-iv. (i-v) The evolution of the *c*-axis orientation through the thickness of the sample, with the creation and annihilation of topological defects. Streamlines represent the in-plane components of the orientation and are coloured according to their x component. They are overlaid on the electron density so that the location of volume defects is also visible (white regions). i) A single trefoil defect (T1 in orange) is present near the open surface. Moving up through the thickness, the trefoil defect T1 shifts to be centred on a nano-void (ii), while maintaining its topology, and then it moves out of the nano-void to the other side (iii). iv) a pair of dislocations with opposite topological charge (comet C in blue and trefoil T2 in green) are created at the open surface. v) The opposite charge pair consisting of the trefoil defect T1 and comet defect C annihilates, and a single trefoil defect T2 remains.

Higher magnifications of slices iv) and v) showing experimental renderings of comet and trefoil topological defects (left) with model representations (right).

With the development of ptychographic X-ray linear dichroic orientation tomography (XL-DOT), we demonstrate its application on a polycrystalline $V_2O_5$ pillar for which we not only map out the intra-granular structure, but also locate and identify nanoscale crystallographic and topological defects within the sample. As well as resolving twin, twist and tilt grain boundaries, we are able to map the presence and spatial evolution of comet and trefoil defects revealing both the creation and annihilation of defect pairs of opposite topological charge.

This capability to non-destructively map the microstructure of materials represents a significant advance in the state-of-the-art where, until now, such information was only available through destructive or lower spatial resolution techniques. XL-DOT takes an important position amongst the current microstructural characterisation techniques covering, in part, spatial resolutions previously reserved for electron diffraction and allows for the non-destructive examination of system-representative sample volumes unrestricted by the size of the probe.[7,10,12–15].

This opens the door to the *operando* mapping of a material's composition and microstructure on the nanoscale. Here, one can envisage not only following the evolution of the microstructure during material synthesis, under annealing conditions, or under mechanical loading, but also mapping the evolution of the microstructure during the cycling of, for example, solid state batteries, fuel cells or heterogeneous catalysts under operational conditions. This ability to identify the role played by the microstructure on material performance will directly inform the improvement of future devices across a wide range of societally relevant industries.

Beyond mapping the microstructure of crystalline systems, the dichroic contrast mechanism can be used to map the orientation of amorphous materials that until now could not be characterised with diffraction-based techniques alone[2,35–37]. This would allow, with nanoscale precision, the 3D characterisation of natural polymers in bone and synthetic polymers in solar cells, as well as secondary phases located in grain boundaries or material inclusions within grains. Moreover, while we focus here on the microstructural characterisation of materials with the probing of electronic anisotropies, the extension to linear magnetic dichroism is a natural progression[38,39], and will allow the characterization of previously inaccessible 3D magnetic topological textures in antiferromagnets[40,41].

Finally, with the use of coherent X-rays, XL-DOT stands to benefit dramatically from increases in coherent flux with the fourth generation of synchrotron radiation sources[42], which will bring the spatial resolution down to the order of ten nanometres, providing a way to map smaller point and line defects, such as dislocations.

**Methods:**

*Materials:* V$_2$O$_5$ was purchased from US Research Nanomaterials. Polystyrene latex spheres, 330 nm in diameter, were purchased from ThermoFisher Scientific.

*Sample Preparation:* The examined pillar was extracted from a sintered millimetre-sized pellet. The latter was prepared from a mixture of nanocrystalline V$_2$O$_5$ and polystyrene spheres (85/15 wt. %). Mortar-and-pestle (10 minutes) were used to homogenize the mixture before the pellet was pressed using a 17 mm die set (3 minutes, 1.2t uniaxial load). To increase the V$_2$O$_5$ grain size, sinter and create the desired porous structure, we heated the pellet to 590ºC for 5 hours, Figure S1. The polycrystalline V$_2$O$_5$ pillar was prepared by mechanically fracturing the sintered pellet, after which a fracture piece was mounted on an OMNY tomography pin[43] using epoxy resin. The pillar was then pre-shaped using a micro-lathe[44], before being reduced in diameter to 6 µm using focused ion-beam (FIB) milling. This pillar was then transferred onto a second OMNY pin. The tip of the second OMNY pin was sharpened using FIB milling prior to transferring the pillar. This was necessary to facilitate tomography measurements with a 30º stage tilt[23]. See Figure S3 for micrographs of the prepared pillar.

*General Material Characterization:* Scanning electron microscopy (SEM) and FIB milling were performed using a Zeiss NVision-40 Dual-Beam FIB. Powder X-ray diffraction (PXRD) measurements of the sample before and after sintering were acquired using a Cu-K alpha radiation source with a step size of 0.02º 2θ, Figure S2[45,46]. The sintered sample consists of α-V$_2$O$_5$ with a grain size > 100 nm.

*Origin of Linear Dichroism in α-V$_2$O$_5$:* V$_2$O$_5$ possesses a layered orthorhombic crystal structure consisting of distorted [VO$_5$] pyramids, shown schematically in Figure 1b. These pyramids tile along the ab plane and are bound with Van der Waals interactions along the c axis. The apical, vanadyl bond, of these pyramids, aligned with the crystallographic c-axis, is shorter (1.57 Å) compared to the bonds on the base of the pyramid (1.87 Å). This shorter bond breaks the symmetry of an otherwise regular square pyramid[47]. To probe the spatial orientation of the apical bond and, in turn, the orientation of entire grains and deviations within them, the energy of the incident X-rays was set to that of the vanadium K pre-edge peak[47]. This peak arises from the V (1s) to V (4p-3d) transition, more specifically to V (3d e$_g$ + 4$_p$) + O 2p$_z$ mixing states, which become accessible as a result of the deviation of the V coordination from the octahedral symmetry. When the apical bond is parallel to the direction of the electric field of the incident X-rays, the interaction is strong as the transition V (1s) to V (4p-3d) is allowed. When the apical bond is instead perpendicular to the incident polarization, the interaction is weaker[27,47]. An illustration of the different absorption strengths that result from the relative orientation between the incident X-ray polarization and the apical bond, known as linear dichroism (LD), is shown in Figure 1b. The X-ray near-edge absorption and phase spectra of V$_2$O$_5$ measured using linear horizontal and vertical polarizations are shown in Figure S4, and a schematic of the layered crystal structure is provided in Gao et. al. (2020)[48].

*Ptychography, Ptychographic X-ray Computed Tomography (PXCT) and Phase Contrast:* Ptychography is a lensless imaging technique in which the phase problem is solved by means of iterative phase retrieval algorithms[25]. By applying ptychography to solve the phase problem at different projection angles, its tomographic extension PXCT[49], is able to retrieve the complex-valued transmissivity of the specimen, providing quantitative tomograms of both phase and amplitude contrast[30]. Both the individual images – or projections – and resulting tomograms obtained using X-ray ptychography are sensitive to changes in the complex valued refractive index, η. The real part of the refractive index decrement, *δ*, corresponds to the phase, while the imaginary part of the refractive index corresponds to the amplitude, $\beta$. The refractive index is fundamentally an expression of the complex atomic scattering factor, $f = f_1 + if_2$. The refractive index is therefore given by:

$$n = 1 - \delta - i\beta = 1 - \frac{r_e}{2\pi}\lambda^2 \sum_k n_{at}^k \left(f_1^k + if_2^k\right), \qquad (S1)$$

with $r_e$ being the classical electron radius and $\lambda$ the illumination wavelength[50,51]. The images and tomograms resulting from measurements performed with incident X-ray energies away from sample-relevant absorption edges can, in the case of tomograms, be converted to quantitative electron density, $n_e$, and absorption index, $\mu$, tomograms[30]. Measurements conducted near sample-relevant absorption edges, i.e. probing specific electronic transitions and the associated increase in the photo absorption cross-section, are subject to anomalous scattering effects[50,51], including dichroism. We make use of this dichroic effect, the dependence of the absorption cross-section on the polarization state, to probe the *c*-axis orientation of vanadium in $V_2O_5$.

Although XL-DOT can be applied with a range of imaging techniques, such as Scanning Transmission X-ray Microscopy (STXM), we have selected X-ray ptychography as the imaging modality, a selection motivated by three factors. (1) PXCT provides quantitative or absolute contrast tomograms, which is ideal for chemical component identification and for the detection of marginal signal variations[11,28]. (2) As a lensless imaging technique, ptychography excels in terms of signal-to-noise ratio (SNR), spatial resolution and dose efficiency (per resolution element) compared to other methods[52–54]. Given its superior SNR, it is ideal for measuring the relatively weak (LD) signal exhibited by $V_2O_5$[28,55]. (3) Ptychography can access the phase component. Phase changes at the vanadium K edge are twice as large as changes in the absorption, so that the retrieved phase projections possess a higher spatial resolution and superior SNR, see Figures S5-6[51]. All data analysis was performed on the *phase component* of the projections and tomograms only.

**Ptychographic Linear Dichroic X-ray Tomography**

*Data Acquisition:* Experiments were carried out at the cSAXS beamline of the Swiss Light Source. The photon energy was selected using a double crystal Si (111) monochromator. The horizontal aperture of slits located 22 m upstream of the sample was set to 20 μm, creating a virtual source point that coherently illuminates a 220 μm diameter Fresnel zone plate with an outermost zone width of 60 nm, and with engineered aberrations designed to improve reconstruction contrast and spatial resolution[31]. Coherent diffraction patterns were acquired using an in-vacuum Eiger 1.5M area detector, with a 75 μm pixel size, placed 5.235 m downstream of the sample inside an evacuated flight tube. Tomography experiments were performed using the positioning instrument described in Holler *et. al*[56].

To map the local orientation of the apical bond within the examined sample volume in 3D, we exploited its LD and acquired 8 equiangular ptychographic tomograms over 180° at 5.469 keV for different illumination polarizations and sample tilts. Specifically, ptychographic tomograms were acquired with a linear horizontal (LH) and linear vertical (LV) polarization of the incident illumination at 0° stage tilt and at 30° stage tilt (sample in grey and pink in top two panels to the right of Figure 1a). Two additional tilts were measured whereby the sample was first rotated by +90° and -90° about the main axis of the pillar, followed by a 30° stage tilt[23]. The latter two tilts are equivalent to tilting towards and away from the beam by 30° (sample in green and blue in bottom two panels to the right of Figure 1a). Examination under different sample tilts and X-ray polarizations is required to have sufficient information for the construction of an orientation tomogram representative of the apical bond orientation in 3D[23,57]. To change the illumination source native horizontal polarization to vertical, we used a 250 μm - thick diamond crystal phase plate inserted into the illumination path upstream of the zone plate (see Figure 1a). The phase plate absorbed approximately 65% of the incident photons[58]. The degree of polarization of the X-rays was determined to be approximately 60% using a polarization analyser setup. The sample tilt was changed using a sample holder insert[23]. To minimize the acquisition time, we utilized an adaptive field-of-view for each group of ptychographic projections. The maximum field-of-view, horizontal × vertical, was ~24 × 25 µm$^2$. The scanning followed a Fermat-spiral pattern[59]. An average step size of 0.8 µm was used for all tomograms. The exposure time per scanning point was 0.1 seconds. 280 projections were acquired per tomogram.

Finally, using the same acquisition parameters, we acquired an off - resonance ptychographic tomogram of the pillar below the absorption edge at 5.4 keV. This tomogram, being insensitive to any dichroic effects, was used for computing the electron density tomogram, and subsequently utilized for compositional analysis[11]. It should be noted that the starting angle and angular spacing of projections was kept constant across all tomograms.

***Ptychographic Image Reconstruction:*** Ptychographic images (or tomographic projections), were reconstructed using the PtychoShelves package[60]. For each reconstruction, a region of 600 × 600 pixels of the detector was used per scanning point, resulting in an image pixel size of 30.91 nm for the pre-edge and 31.29 nm for the below-edge tomogram. Reconstructions were obtained with 200 iterations of the difference map (DM) algorithm[61] followed by 300 iterations of maximum likelihood (ML) refinement[62].

***Preprocessing of Projections:*** Prior to any tomogram reconstructions, we (1) resampled all projections to a pixel size of 30.91 nm using Fourier interpolation, (2) extracted the phase from the reconstructed projections, removed constant and linear phase components, and spatially aligned the projections using a tomographic consistency approach[29], and (3) aligned all projections to a common pillar orientation. As a last step, the different orientations at which projections were measured were characterized by a 3D rotation matrix[23] which was input into a specially developed reconstruction code (see XL-DOT Reconstruction below). It should be noted that, due to the sample tilt and the fixed vertical field-of-view of the 2D projections, the 3D volume that is commonly sampled in all orientations, and used in the subsequent analysis and visualization, is reduced. (4) Lastly, to isolate the dichroic component from the isotropic electron density contribution, the LV projection was subtracted from the LH projection. The resulting set of projections were used in the reconstruction of the XL-DOT dataset as discussed further below.

***Ptychographic Tomogram Reconstruction:*** The ptychographic tomogram, acquired with the X-ray energy tuned to below the absorption edge, was reconstructed using a modified filtered back-projection algorithm (FBP)[63]. This off-resonance phase tomogram was used to derive the electron density tomogram, which was then used for material component identification[11,30].

***XL-DOT Reconstruction:*** A gradient-based iterative reconstruction algorithm was developed to reconstruct the crystallographic *c*-axis orientation. A schematic of the reconstruction process is shown in Figure S7. The process starts with the creation of a 3D starting, random guess of the sample. Using the sample-illumination interaction relationship Equation (2), a set of projections is simulated. These projections are then compared to the measured set of projections, and their difference is used to compute a gradient to iteratively correct the initial guess.

The interaction between the electric field of the incident linearly polarized X-rays, $\vec{E}$, and the orientation of the apical vanadyl bond, $\vec{a}$, can be described with the formulation given in Equation (2). Here, $f$ is the total scattering factor, which contains the isotropic charge contribution, $f_0$, and the LD contribution, $(\vec{E} \cdot \vec{a})^2$, with a pre-factor $f_{lin}$ that depends on the electronic transition under resonance.

$$f = f_0 + f_{lin}(\vec{E} \cdot \vec{a})^2 \qquad (2)$$

Adhering to the experimental geometry (Figure 1a); using X-rays with a LH polarization parallel to the x-axis, and denoting an arbitrary polarization angle as $\varphi$, where $\varphi = 0°$ is LH and $\varphi = 90°$ is LV polarization, the tomographic rotation and tilting of the sample can be quantitatively represented by the 3D rotation matrix **R**. In transmission, the measured projection can then be described by the integral given in Equation (3). Index summation notation is used to give the rotation of the relevant components of the orientation, $a_j$. The integration is evaluated along the X-ray propagation direction, the z axis.

$$P(x, y) = \int f_0(\mathbf{R}\vec{r}) + f_{lin} \left[ R_{1j} a_j (\mathbf{R}\vec{r}) \cos \varphi + R_{2j} a_j (\mathbf{R}\vec{r}) \sin \varphi \right]^2 dz \qquad (3)$$

Knowing the form of the interaction, the reconstruction algorithm was formulated by generating a guess structure, from which projections were simulated at the same orientations that the sample was measured. These simulated projections, $\hat{P}$, were then compared with the corresponding measured projections, $P$. Their square difference was used to define an error metric, $\epsilon$, quantifying how well the guess could reproduce the measured projections, given by

$$\epsilon = \sum_{m,x,y} \left[ \hat{P}^m(x, y) - P^m(x, y) \right]^2 \qquad (4)$$

Here, $m$ represents the projection index. The error metric was reduced using gradient descent, therefore improving the ability of the guess structure to represent the internal *c*-axis orientation of the measured sample. By differentiating the error metric in Equation (4) with respect to each component, we obtain the following analytical expression for calculating the gradient:

$$\frac{\partial \epsilon}{\partial a_k} = 2 f_{lin} \sum_{x,y} \left[ \hat{P}^m(x, y) - P^m(x, y) \right] \left[ R_{1j} a_j \cos \varphi + R_{2j} a_j \sin \varphi \right] (R_{1k} \cos \varphi + R_{2k} \sin \varphi) \qquad (5)$$

The gradient was evaluated and applied to the guess structure at every iteration. As the iterative gradient descent reconstruction is prone to converging at local minima, 40 individual reconstructions were performed using different random, non-zero initial conditions. The individual reconstructions are combined by averaging all components to obtain a final reconstruction. The difference in the angular orientations between the individual reconstructions and the final, averaged reconstruction was used to evaluate the standard deviation of the orientation, which is an estimate of the uncertainty in orientation.

Interestingly, using Equation (3) it can be shown that LV polarization ($\varphi = 90°$) projection measurements evaluate to

$$P(x,y) = \int \left(f_0(R\vec{r}) + f_{lin}[\boldsymbol{a_y}(R\vec{r})]^2\right) dz \qquad (6)$$

Since there are no vector rotations in this expression, it is equivalent to probing a scalar consisting of two components: the isotropic charge background, $f_0$, and the (out-of-plane) $a_y^2$ component. This can be reconstructed with conventional tomography and gives contrast between grains that are in-plane (xy-plane) and out-of-plane oriented. This contrast was used for additional validation of the final reconstruction, as shown in Figure S12.

*Dose Estimation:* The total deposited dose over the duration of the experiment and the entire volume of the $V_2O_5$ pillar is approximately $10^9$ Gy. This estimate is based on the sample's mass density and the average flux density per projection[64].

*Spatial Resolution:* Spatial resolution estimates of projections and tomograms were obtained using Fourier ring correlation (FRC) and Fourier shell correlation (FSC), respectively[65].

To evaluate the spatial resolution of the acquired projections, we acquired two projections under identical conditions i.e. at the same rotation angle, calculated the correlation between these two images in the Fourier domain and estimated the spatial resolution based on the intersection with a one-bit threshold, see Figure S6. This gives spatial resolutions close to the pixel limit of 30.91 nm and 31.29 nm for the on-resonance (5.469 keV) and off-resonance (5.4 keV) measured projections, respectively.

To evaluate the spatial resolution of the electron density tomogram acquired below the absorption edge, we divided the entire dataset in half and reconstructed two independent tomograms, Figure S10. This gives a 3D spatial resolution of 44 nm.

To evaluate the spatial resolution of the orientation vector field, the corresponding dataset was similarly split in half and two tomograms of the orientation vector field were calculated. Using FSC, we calculated spatial resolution estimates for each of the orientation scalar components ($LD_x$, $LD_y$, $LD_z$), as shown in Figure S8, providing a lower bound for their spatial resolution of 84 nm, 45 nm, and 89 nm, respectively. In addition, we measured edge profiles across sharp features such as 90° grain boundaries, which revealed a maximum edge sharpness of 40 nm, with an average edge sharpness of 73 nm, which we take as the spatial resolution of the orientation tomogram.

*Measurement Error Estimation:* To estimate the voxel level electron density uncertainty, we calculated the standard deviation (σ) of the electron density in a region of air surrounding the imaged pillar. The average electron density and uncertainty was calculated to be $0.004 \pm 0.007$ Å$^{-3}$.

To estimate the uncertainty in the detected LD, i.e., spatial variations in the pre-edge peak intensity, we first reconstructed a LV phase tomogram and a LH polarized phase tomogram of the sample at a fixed sample tilt, and then subtracted them from each other. We then isolated a region of air and calculated the standard deviation in the phase shift associated with the voxels in this region. Based on this procedure, the average measured phase shift with standard deviation was $5.8 \times 10^{-6} \pm 1.3 \times 10^{-4}$ rad, which corresponds to a refractive index decrement, $\delta$, of $1.9 \times 10^{-7} \pm 1.3 \times 10^{-9}$.

To estimate the error in the determined orientation, we isolated an elongated grain with a volume of 0.85 μm$^3$ and long-edge length of 3.2 μm that displayed the least variance in electron density and $V_2O_5$ orientation, i.e. which is assumed to be single crystal, and calculated the standard deviation (σ) in orientation to be ± 10° for azimuth (xy-plane angles) and ± 8° for elevation (out-of-plane angles), Figure S11.

## Data Analysis

Dichroic tomogram analysis was performed using in-house developed MATLAB routines, Paraview and Avizo. To avoid artefacts due to the FIB sample preparation, we defined a mask which excluded the outermost 90 nm of the sample cylinder from orientation and electron density volume analysis.

***Component Identification and Isolation:*** Materials were identified by comparing the tabulated electron densities of the known sample and reference components, listed in Table S1, with the PXCT measured electron densities. Shown in Figure S9 is a volume rendering and a horizontal cut slice through the electron density tomogram with the corresponding electron density histogram. The $V_2O_5$ volume was isolated using threshold segmentation with a lower-bound of 0.74 Å$^{-3}$ and an upper-bound of 0.90 Å$^{-3}$.

***Microstructural Analysis of $V_2O_5$ domains:*** To isolate the $V_2O_5$ grains and facilitate a correlation between orientation and electron density, we applied the above-defined threshold mask (electron densities between 0.74 Å$^{-3}$ and 0.90 Å$^{-3}$) to the orientation tomogram. To identify and characterize individual $V_2O_5$ grains we downsampled the masked XL-DOT reconstruction by a factor of three (transforming a group of 3×3 voxels and into 1 voxel with an average intensity value of the same size), thus reducing the sensitivity to intra-granular variations. Segmentation was then performed by separating regions along high-angle grain boundaries (HAGBs), displaying a *c*-axis orientation difference larger than 10. Following segmentation, we then calculated the volume of these grains, their mean diameter, and sphericity[66]. Shown in Figure S13 are the corresponding distributions and correlations of the segmented grains.

**Acknowledgements**

**General:** Tomography experiments were performed at the coherent small-angle X-ray scattering (cSAXS) beamline of the Swiss Light Source (SLS), Paul Scherrer Institute (PSI). We thank, X. Donath and P. Zimmerman for technical support. Electron microscopy work was performed at the Scientific Centre for Optical and Electron Microscopy (ScopeM) at ETH Zurich and the Electron Microscopy Facility at PSI, with the assistance of E. A. Mueller Gubler, J. Reuteler and A. G. Bitterman. We further would like thank Ian Robinson for the insightful discussion.

**Funding:** The research leading to these results has received funding from the Swiss National Science Foundation (SNSF), with project numbers 200021_192162 (A.A.), 200021_196898, & PZ00P2_179886 (Z.G. & J.I.), the Max Planck Society Lise Meitner Excellence Program (C.D.) and the EU's Horizon 2020 research and innovation program under the Marie Sklodowska-Curie grant agreement No. 884104 (PSI-FELLOW-III-3i) (C.A.).

**Author Contributions:** J.I., C.D. and M.G.S. conceived the study. V.S. and M.H. designed, setup and aligned the polarization stage. A.A., M.H., M.G.S., C.A., C.D. and J.I. performed PXCT experiments. A.A. and C.D. developed the orientation tomogram reconstruction algorithm. A.A. performed PXCT and X-ray linear dichroic orientation tomography reconstructions with support from C.D. A.A. analysed the data with the help of C.D., M.G.S. and J.I. P.Z and J.I. prepared the sample. A.A., C.D. and J.I. wrote the manuscript. L.H., M.G.S., M.H. and V.S. contributed to discussions. All authors helped improve and approved the manuscript.

**Competing Interests:** The authors declare no financial, ethical, or other competing interests concerning the contents of the presented manuscript.

**Data Availability:** Information needed to evaluate the presented conclusions are provided in the manuscript and the Supplementary Information. The raw data and reconstructed projections can be accessed at the following address, doi.xyz (*to be populated during final manuscript revisions*) or obtained from the corresponding authors.

**Code Availability:** General acquisition and reconstruction codes can be accessed at https://www.psi.ch/en/sls/csaxs/software. The orientation tomogram reconstruction method and code will be published in a separate technical paper. A version of the code utilized here can be obtained from the corresponding authors.

**Supplementary Information**

**This Supplementary Information Includes:**

- Supplementary Note 1
- Supplementary Figure 1-13
- Supplementary Table 1
- Supplementary Movies 1-3

**Supplementary Note 1: X-ray Linear Dichroism as Reflected in the Pre-edge Peak Intensity of $V_2O_5$:** X-ray linear dichroism (XLD) is a measure of the local changes of a material's refractive index as a function of incident beam polarization and energy.[27] Specifically, XLD, refers to the anisotropic absorption of linearly polarized electromagnetic radiation due to the orientation of a sample feature relative to the electric field vector of the radiation [67,68]. XLD contrast is present in the X-ray absorption near-edge structure (XANES) region for photon energies in the pre-edge peak, providing sensitivity to the coordination geometry of the probed chemical element. The polarisation and sample orientation-dependent absorption cross-section in XLD measurements allows for the determination of the spatial orientation of the probed chemical bonds. Multiple measurements at the pre-edge peak energy, in this case 5.469 eV as a function of the linear polarization state of the illumination, allows the determination of the orientation of the probed chemical bonds[67,68] The intensity, $I$, of near-edge features, including the pre-peak intensity, depends as explained in Fu et. al. (2005)[68] on the angle, $\varphi$, between the electric field vector (in other words, the linear polarization vector of the illumination) and the angle-integrated absorption cross-section, A (Equation S1).

$$I \approx A \cos^2 \varphi \qquad (S1)$$

A pre-requisite for any linear dichroism is the presence of an anisotropic "bonding environment". The $V_2O_5$ examined here satisfies this pre-requisite due to its layered orthorhombic crystal structure. XLD and the resulting analytical capabilities have been utilized in a microscopy context before, i.e., X-ray linear dichroism microscopy[27,28,67,69–71], to provide a 2D spatially-resolved analysis tool. The reader is directed to the initial work of Ade[67] and the more recent works of Gilbert[55,71–73] and Collins[69,70,74]. XLD microscopy and tomography are particularly suited for the characterization of materials with a net-orientation or anisotropy. Polymer arrangements[35], most non-cubic and some cubic crystalline materials, for example, possess such an anisotropy over extended volumes.

**Supplementary Figure S1: Schematic Illustrating the Preparation of the Sintered Polycrystalline Vanadium Pentoxide Sample.** To increase the $V_2O_5$ grain size and to assemble a stable porous structure from the initially loose powder, a mixture of $V_2O_5$ and polystyrene spheres was pressed into pellet shape with a 1.2 t axial load. The resulting pellet was then heated to 590ºC for 5 hours, to increase the grain size, sinter the structure and decompose the polystyrene as much as possible, allowing the structure to condense further.

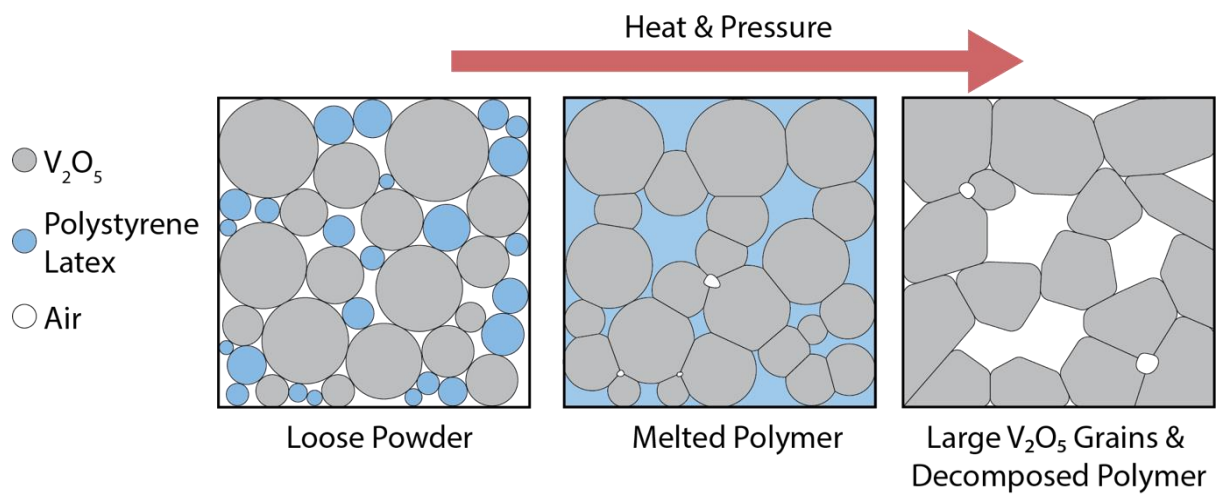

**Supplementary Figure S2: Powder X-ray Diffraction of Vanadium Pentoxide.** Shown in black is PXRD data collected from $V_2O_5$ powder. Shown in orange is data collected from the sintered $V_2O_5$ pellet, prior to being manufactured into the sample pillar. All reflections can be assigned to orthorhombic $V_2O_5$[46]. Rietveld refinement[75] using an anisotropic size-strain model[76] suggests the average crystal in the sintered pellet to be asymmetric in shape, with a short axis coherence length of > 300 nm . Instrument broadening was taken into account. The shift observed after sintering is likely to be due to lattice contraction.

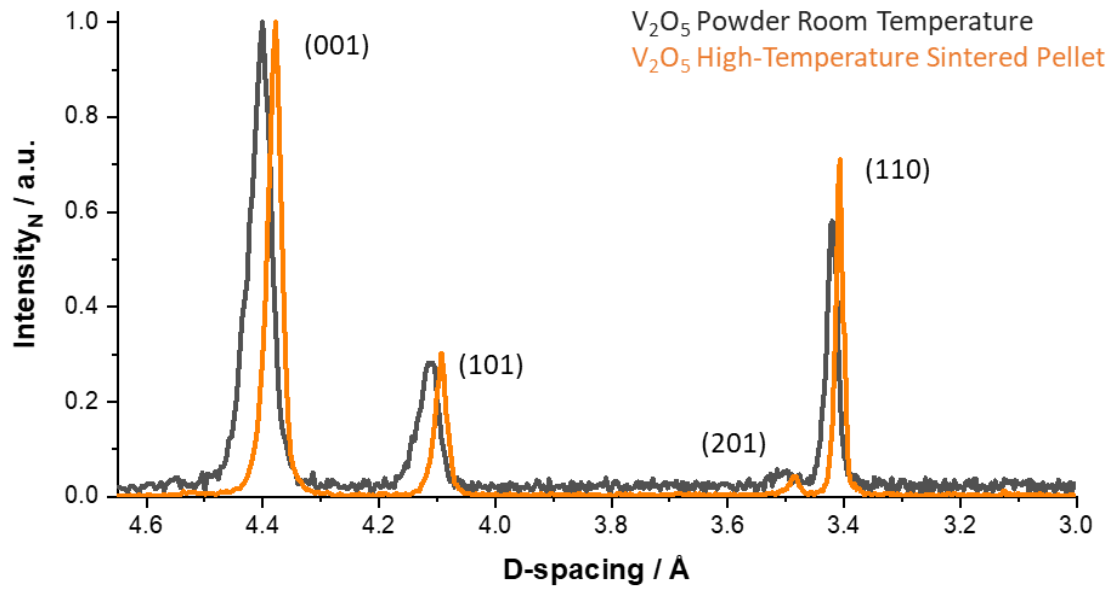

**Supplementary Figure S3: Images of the Examined V$_2$O$_5$ Pillar.** Shown is the progression of sample preparation including (a) a photograph of the fractured sintered pellet, and subsequent Scanning Electron Microscopy (SEM) images of (b) the micro-lathe pre-shaped pillar, (c) the FIB-milling reduced pillar, (d) the tomography-pin transfer and (e) the final mounted sample pillar. Scale bars are 1000 µm (a) and 20 µm (b-e).

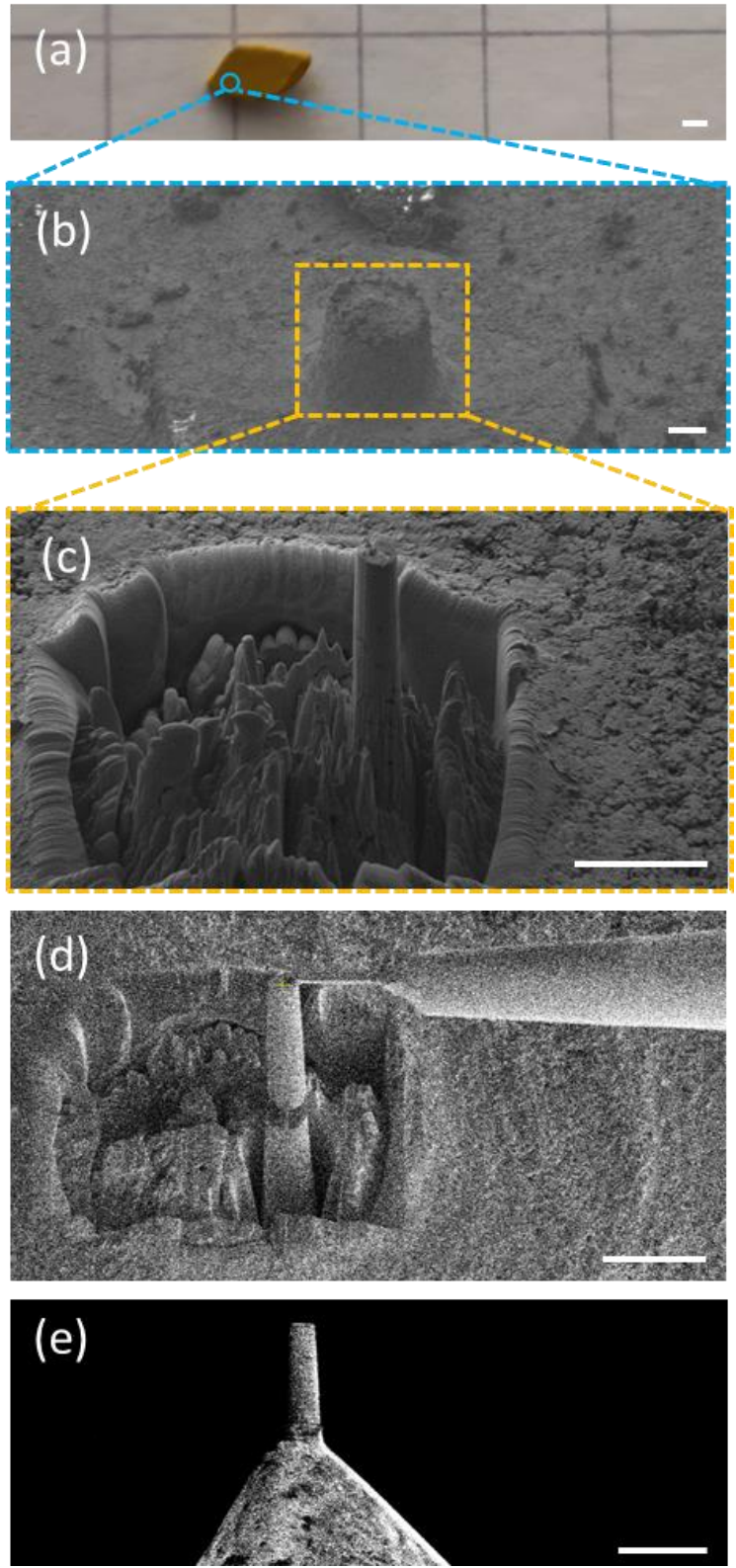

**Supplementary Figure S4: X-ray transmission near-edge spectra of $V_2O_5$.** Spectroscopic data was acquired using both LH and LV X-ray polarizations, showing both the energy and polarisation dependency of the X-ray linear dichroism of $V_2O_5$. Grey lines indicate the energies at which tomographic projection measurements were made for the off-resonance and on-resonance tomograms. On resonance, the linear dichroism effect can be observed in both amplitude and phase, as the sample-illumination interaction strength changes with polarization. The optimal LD constrast in the phase and amplitude occur at different energies. The optimal energy for maximizing phase contrast was chosen to be 5.469 keV due to the higher resolution of individual projections. The projections measured at this energy were used for the reconstruction of the orientation tomogram. The tomographic projections measured at the indicated off-resonance energy (5.4 keV) were used for the reconstruction of the electron density. As this energy was sufficiently far from the pre-edge peak, which exhibits LD, we could safely assume that only the electron density contributed to the scattering factor.

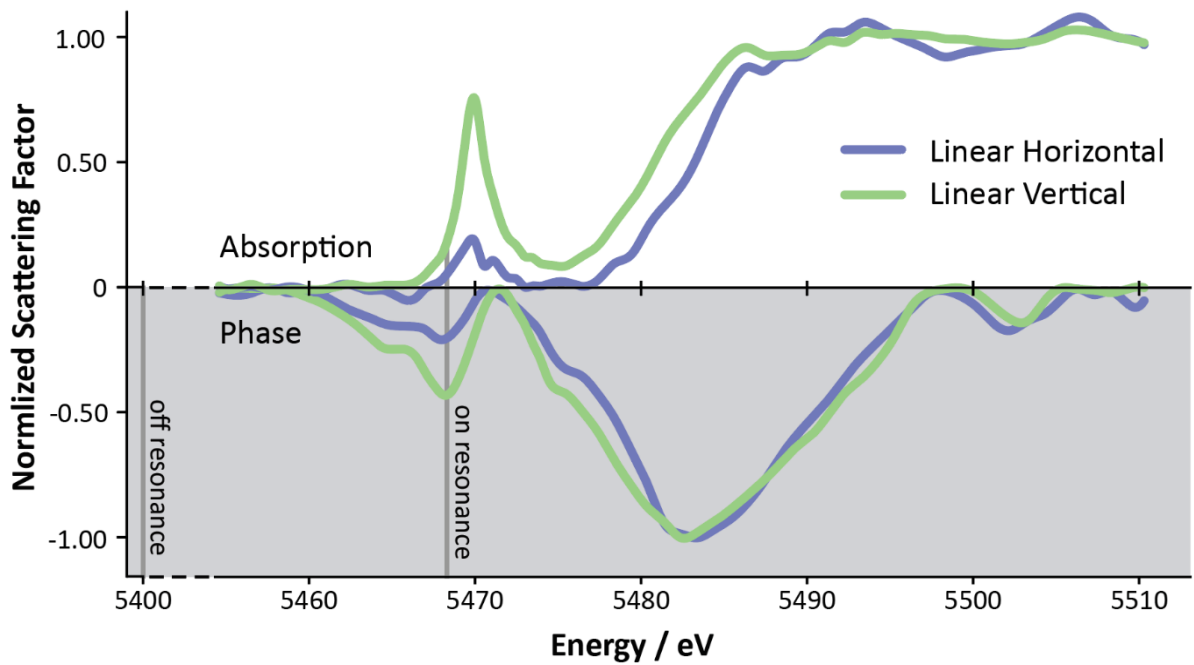

**Supplementary Figure S5: Comparison of Ptychographic Image Reconstructions.** Typical reconstructed X-ray ptychography images (tomographic projections). Comparison of amplitude (top, a-e) and phase (bottom, f-j) reconstructions. (a, g) Reconstructions acquired at 5.400 keV, i.e., below the absorption edge and in the absence of vanadium linear dichroic contrast. Projections acquired at the vanadium pre-peak (5.469 keV) using a linear vertical (left) and linear horizontal (right) polarized illumination are shown in (b-c,g-h). The isolated LD signal is shown as the difference between LH and LV projections (d,i) and the difference between on resonance and off resonance projections for the same polarization (e,j).

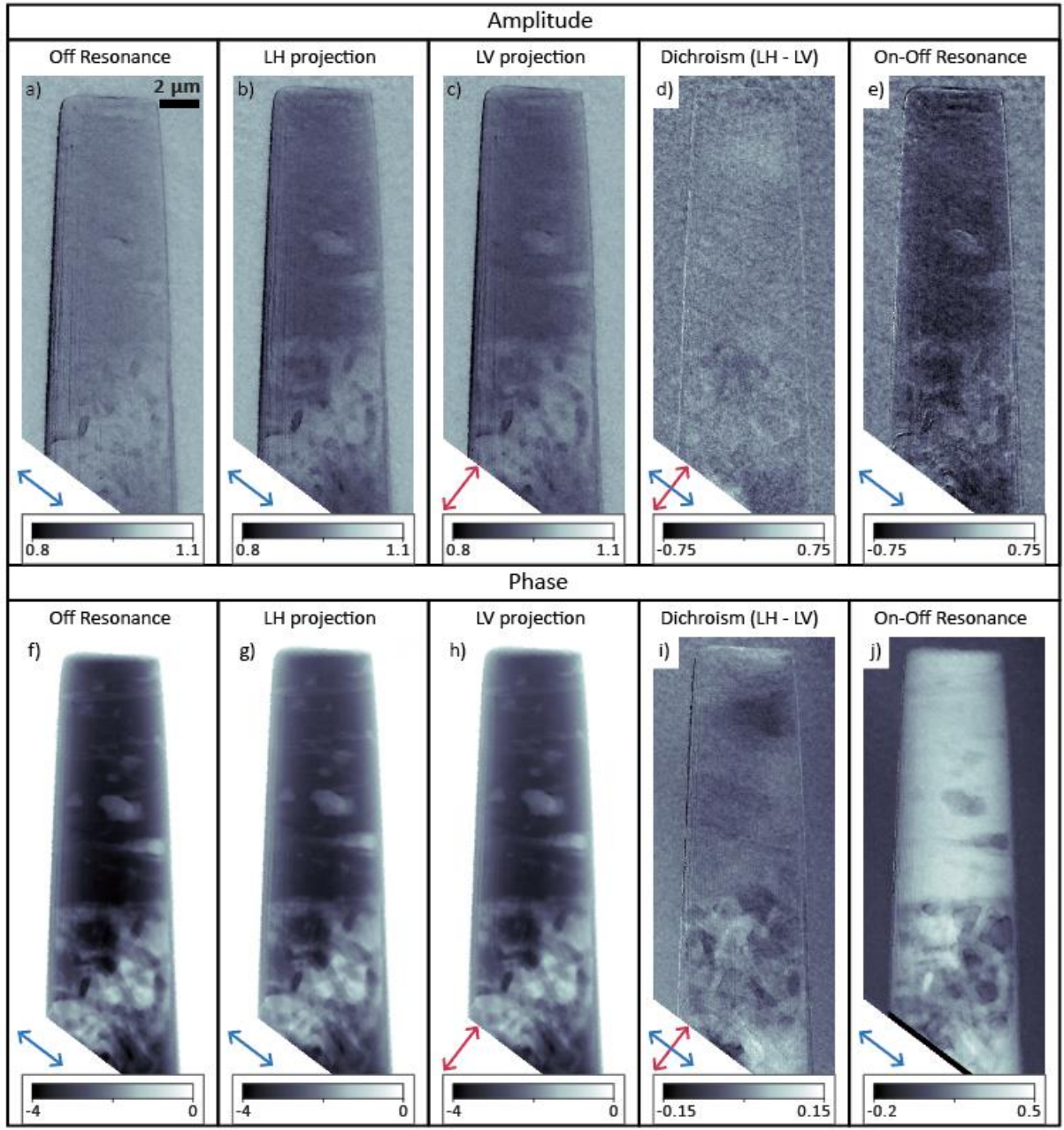

**Supplementary Figure S6: Spatial Resolution of Ptychographic Image Reconstructions.** Fourier ring correlation (FRCs) curves of phase image reconstructions. Correlation curves for (a) projections acquired at the absorption edge (at 5.469 keV) using linear horizontal polarization and (b) below the absorption edge (at 5.400 keV). The former projections are the sum of electron density and linear dichroic contrast, while the latter projections comprise only electron density contrast. The half-pitch resolution is given by the intersection of the correlation curve with the 1-bit threshold. In this case, no intersection implies single pixel resolution. The pixel size is 30.91 nm in (a) and 31.29 nm in (b).

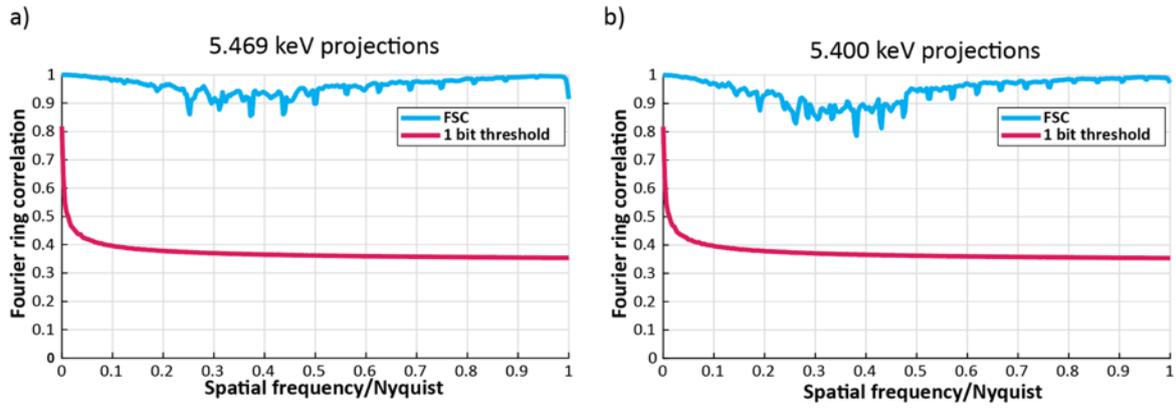

**Supplementary Figure S7: Schematic of the XL-DOT Reconstruction Process.** During the experiment, the sample (i) is used to acquire projections (ii) containing linear dichroic contrast at multiple angles and/or linear polarisation states. The collection of these projections (iii) constitutes the dataset necessary for the subsequent reconstruction process. The reconstruction algorithm begins with an initial guess of the internal structure of the sample (a). For each experimentally obtained projection, a corresponding projection is simulated (b), using the same measurement parameters as in the experiment. The stack of simulated projections (c) and the experimentally obtained projections (iii) are then used to evaluate the error metric (d) according to Equation (4). This metric quantifies how well the guess structure can replicate the experimentally obtained dataset. The objective is to correct the guess structure, (a), such that the error metric is reduced, indicating a closer match between the guess and the internal structure of the measured sample. This is achieved through the implementation of gradient descent. Having calculated the analytical gradient (e) according to Equation (5), a corresponding correction (f) – conventionally a scalar multiple of the gradient – is applied to the test object. This ensures a reduction in the error metric. The iterative repetition of this process continues for a set number of iterations or until a convergence criterion is met, ultimately yielding the XL-DOT result.

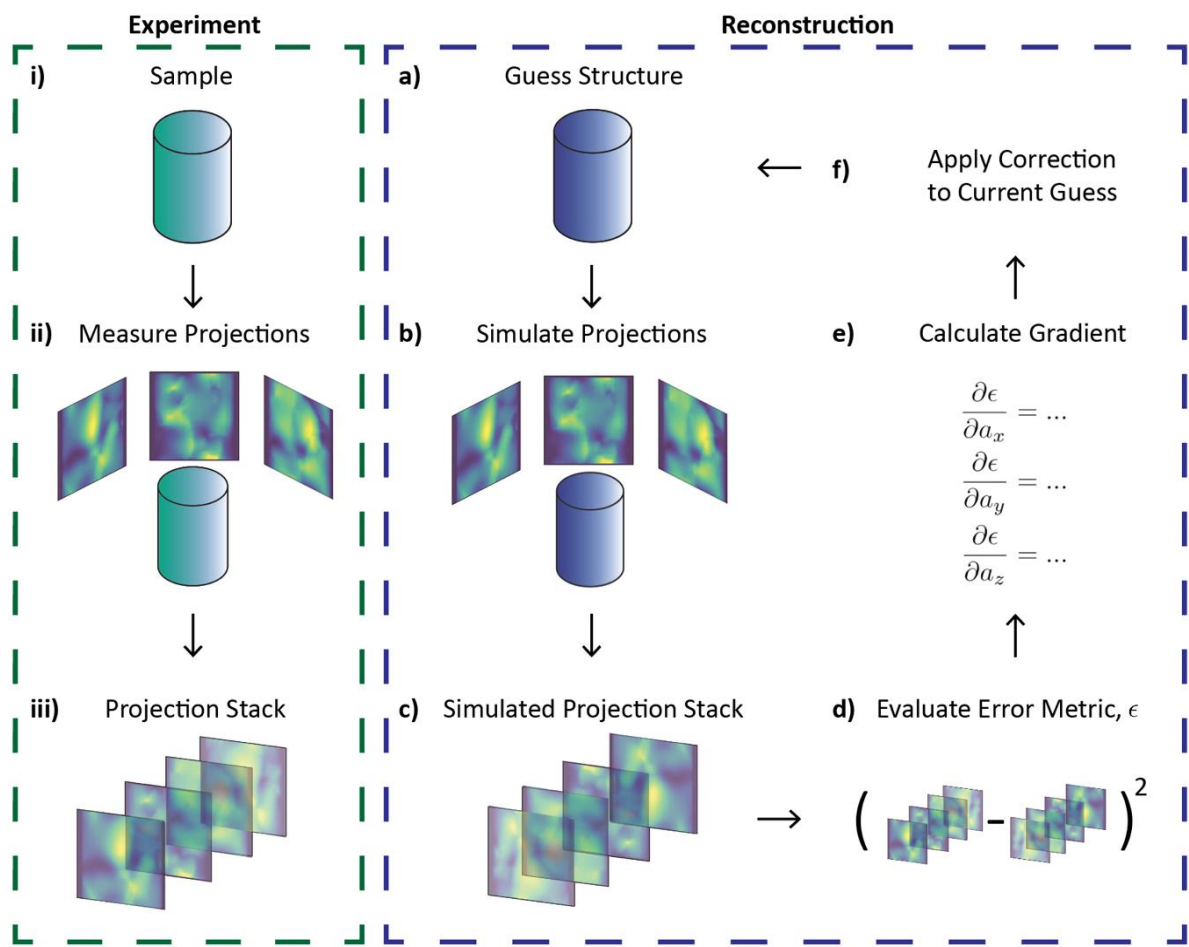

**Supplementary Figure S8: Spatial Resolution of the XL-DOT Reconstruction.** a-d) Estimating resolution from the ability to resolve sharp features. a) Orientation view of chosen sharp boundary (running from top to bottom in the middle of the image) at which the resolution was evaluated. b) Isolation of the out-of-plane component and c) magnified image of the boundary. For each horizontal line of pixels across the boundary in c), a line profile was obtained from which the 10%-90% edge criteria were calculated, with the corresponding value shown on the plot to the right of the image. The red dashed line indicates the average resolution of 73 nm. d) Example of one such line profile, taken from the dashed grey line in c). The edge jump, Δ, is the largest difference in elevation across the boundary, and the resolution of the XL-DOT reconstruction is given by the difference between the 10%Δ and 90%Δ distances. Also provided are Fourier shell correlation (FSC) curves (e-g) of each orientation scalar component (x, y, z), respectively. The half-pitch resolution is given by the intersection of the correlation curves with the half-bit threshold and provided in the figure. The voxel size is 30.91 nm.

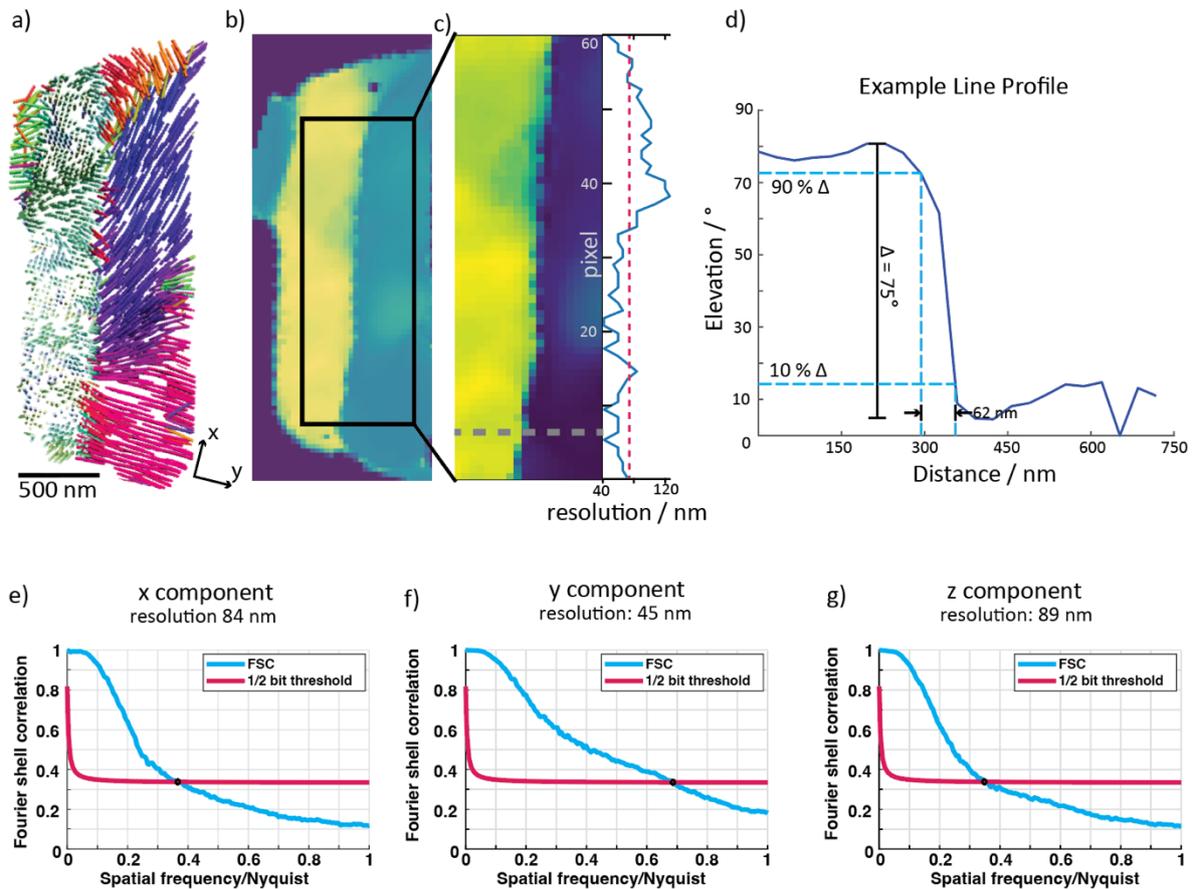

**Supplementary Figure S9: Electron Density Tomogram of the Examined Sample Pillar.** (a) A volume rendering of the acquired electron density tomogram. The voxel volume is $(30.91 \text{ nm})^3$. The red box indicates the volume considered for microstructural examination. (b) Electron density histogram of the volume shown in (a). The expected electron densities of the polymer, defect-rich $V_2O_5$ and pristine $V_2O_5$ are indicated with markers (i), (ii) and (iii), respectively. Depicted by the horizontal bar are the upper- and lower-bounds used for the isolation of $V_2O_5$. Regions of polymer (i) and the defect-rich $V_2O_5$ (ii) are also indicated in panel (c), a slice through the tomogram.

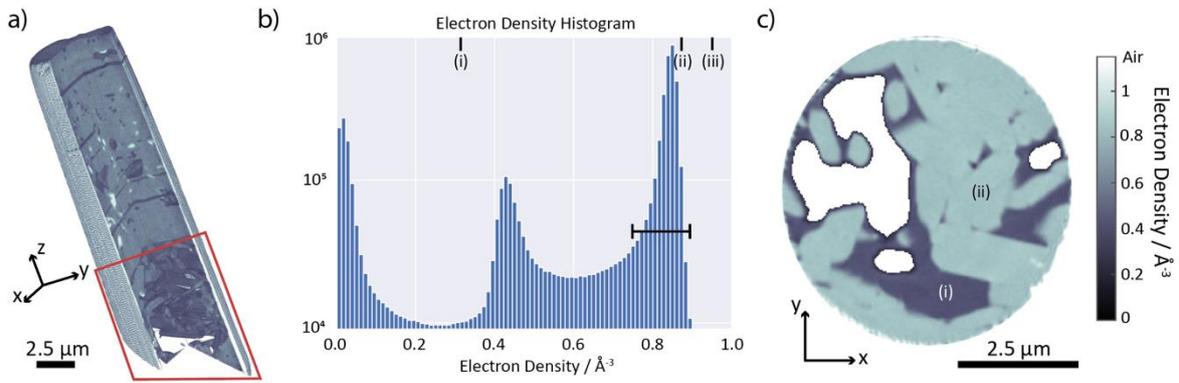

**Supplementary Figure S10: Spatial Resolution of the Electron Density Tomogram.** Fourier shell correlation (FSC) curve of the electron density tomogram derived from the phase projections acquired with the X-rays tuned to an energy below the absorption edge (5.400 keV). The spatial resolution is given by the intersection of the correlation curve with the half-bit threshold. The voxel size is 31.29 nm, and the spatial resolution is 44 nm.

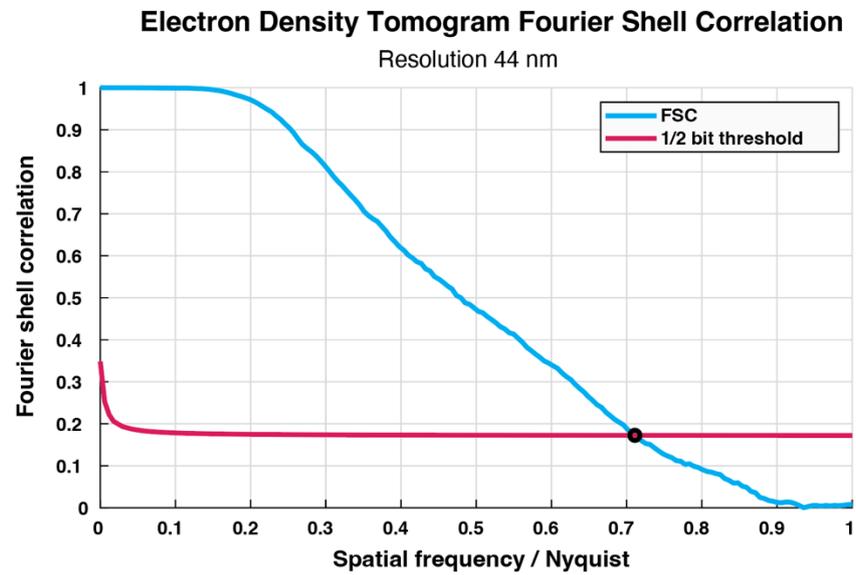

**Supplementary Figure S11: Error in the Determination of the Orientation.** Volume rendering of an isolated $V_2O_5$ grain, exhibiting minimal variations in (a) c-axis orientation and (b) electron density, so that we can assume that this is a single crystal grain. (c) Histogram showing the voxel-level measured orientation spread in this region. From this we determine the following parameters: mean electron density: 0.84 Å$^{-3}$, standard deviation of electron density: 0.024 Å$^{-3}$, Length: 3.2 μm, Azimuthal angle (in-plane, xy-angle): $-70 \pm 10°$, Elevation (out-of-plane angle): $-24 \pm 8°$.

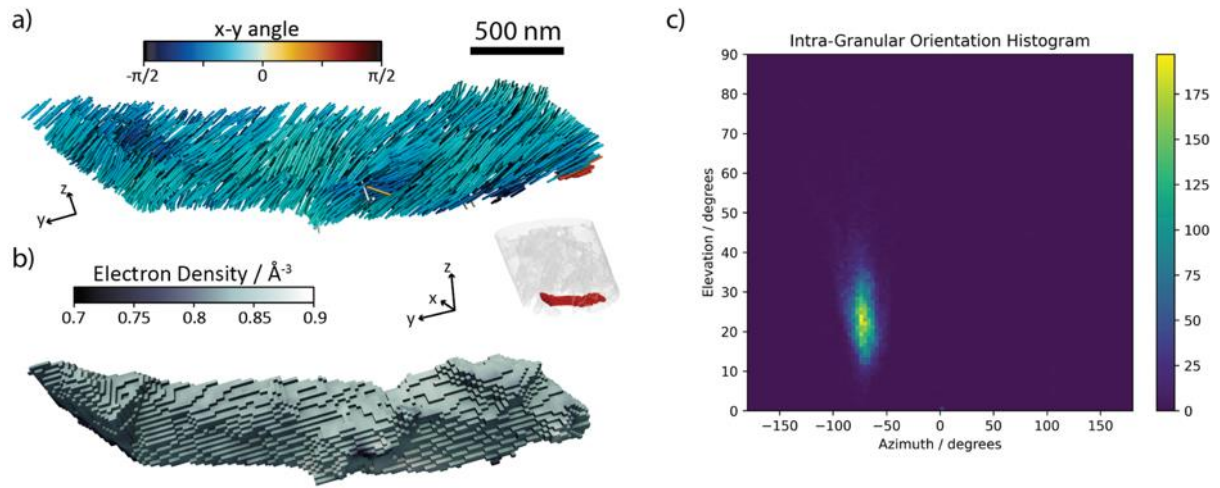

**Supplementary Figure S12: Validation and Comparison of XL-DOT Reconstructions.** (a) Comparison of reconstructed orientation tomograms obtained from the average of 5, 10, 40 reconstructions from different random-valued starting guesses. Provided are virtual cuts through the averaged tomogram. The voxel size is 30.91 nm and figures share a common greyscale as indicated. The same cut, but this time of the scalar reconstruction of the projections obtained using linear vertical polarisation, which is used for validation, is shown on the right for comparison. The contrast of the scalar reference originates from the out-of-plane component of the *c*-axis orientation according to Equation (6). (b) Graph of the angular uncertainty versus the number of averaged reconstructions. The inset is a magnification of the angular uncertainty between 15 and 40 averages, emphasizing the small improvement in angular uncertainty (~0.25˚). As a compromise between computational and time constraints, and small improvements to the reconstruction, the final reconstruction was calculated from an average of 40 reconstructions.

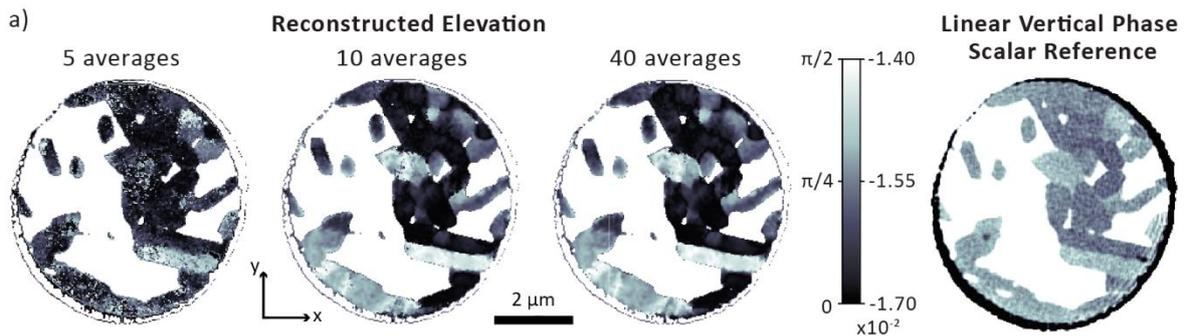

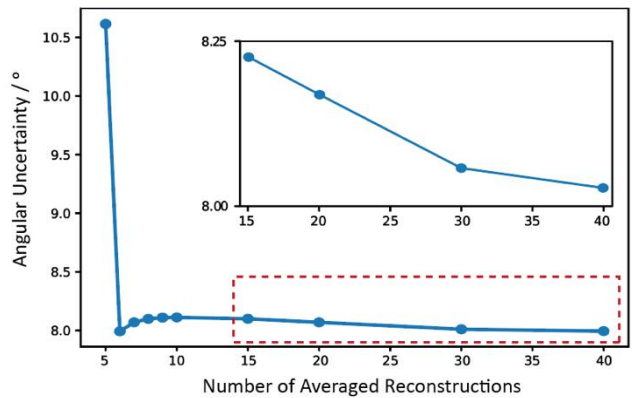

**Supplementary Figure S13: Inter-granular Characterization of $V_2O_5$ Domains.** Histograms showing the correlation between grain volume, grain average electron density, mean length, and sphericity (where 0 and 1 represent an infinite rod and a perfect sphere, respectively). The occurrence frequency histograms are given on the diagonal, with the blue lines corresponding to the kernel density. The upper-right plots (within the orange boundary) are the bi-variate histograms corresponding to the two properties labelled on the vertical and horizontal axes. In the lower-left (within the red boundary), the kernel density estimate (KDE) plots are shown, with the Spearman correlation coefficients between the corresponding grain properties displayed.

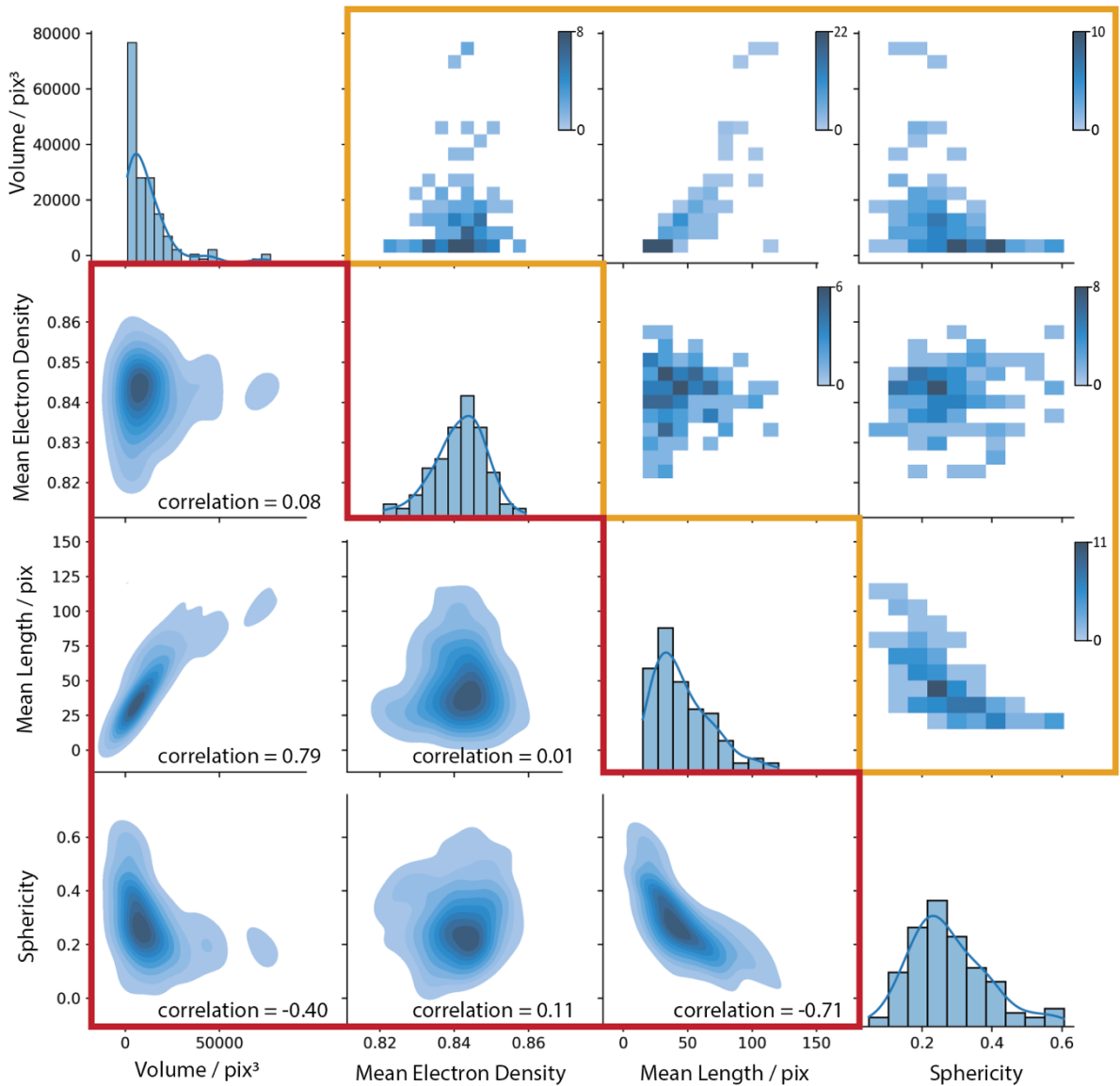

**Movie S1: Electron Density Tomogram.** Animated volume rendering of the acquired electron density tomogram of the sample pillar, showing the morphology of the pillar.

**Movie S2: XL-DOT Reconstruction.** Animated cut slices through the acquired XL-DOT reconstruction revealing the microstructural make-up of the examined sample pillar, and their corresponding electron density slices.

**Movie S3: Topological Defect.** Animated renderings of cut slices in the vicinity of the topological defects and nanovoids, showing the change in $c$-axis orientation.

**Supplementary Table S1: Electron Densities of Reference Components.** Electron densities of relevant materials, including other oxides of Vanadium that $V_2O_5$ could decompose to. Polystyrene polymer was used in the sintering, leading to the porous morphology of the pillar after decomposition. Gallium is also included so that Ga implantation from the FIB processing can be identified and excluded from the analysis. Electron densities were calculated using tabulated molecular weight and mass density values[30].

| Compound | Electron Density / Å$^{-3}$ |
| --- | --- |
| $V_2O_3$ | 1.37 |
| $VO_2$ | 1.29 |
| $V_2O_5$ | 0.96 |
| $V_2O_5$ (Oxygen Vacancy-Rich) | ~0.88 |
| Air / Pores | 0.003 |
| Polystyrene (~$C_8H_8$) | 0.31 |
| Ga | 1.6 |